%% file: lattice2021-arxiv.tex
\documentclass[a4paper,11pt]{article}
\usepackage{pos}
\usepackage{wrapfig}
\usepackage{bbm}
\usepackage{subcaption}
\newcommand{\I}{\mathrm{i}}

\title{$I=1$ $\pi$-$\pi$ scattering at the physical point}

\author*[a]{Srijit\ Paul}
\author[b]{Andrew D.\ Hanlon}
\author[c]{Ben H\"orz}
\author[d]{Daniel Mohler}
\author[e]{Colin Morningstar}
\author[a,f]{Hartmut Wittig}

\affiliation[a]{Institut f\"ur Kernphysik, Johannes Gutenberg-Universit\"at Mainz, \\
Johann-Joachim-Becher-Weg 45, D 55128 Mainz, Germany}
\affiliation[b]{Physics Department, Brookhaven National Laboratory, Upton, New York 11973, USA}
\affiliation[c]{Nuclear Science Division, Lawrence Berkeley National Laboratory, Berkeley, CA, USA 94720}
\affiliation[d]{Helmholtzzentrum für Schwerionenforschung (GSI), Planckstrasse 1, 64291 Darmstadt, Germany}
\affiliation[e]{Department of Physics, Carnegie Mellon University,  Pittsburgh, PA 15213, USA}
\affiliation[f]{PRISMA$^+$ Cluster of Excellence, Johannes Gutenberg-Universit\"at Mainz, \\
Staudingerweg 9, 55128 Mainz, Germany, }
\note{MITP-21-071}

\emailAdd{spaul@uni-mainz.de}

\abstract{We present a preliminary analysis of $I=1$ $\pi\,\pi$ scattering at the physical point. We make use of the stochastic variant of the distillation framework (also known as sLapH) to compute the relevant two-point correlation matrices using a basis of single and multihadron interpolating operators to estimate the low energy spectra. We perform the L\"uscher analysis to determine the scattering phase shift which is finding good agreement with the experimentally obtained phase shifts.}

\FullConference{%
 The 38th International Symposium on Lattice Field Theory, LATTICE2021
  26th-30th July, 2021
  Zoom/Gather@Massachusetts Institute of Technology
}

\input{macro.tex}

\begin{document}
\maketitle

\section{Introduction}
In recent times, with the announcement of the combined results from the $(g-2)_\mu$ experiments at Brookhaven~\cite{Muong-2:2006rrc} and Fermilab~\cite{Muong-2:2021ojo, Muong-2:2021vma} exposing a $4.2\sigma$ tension between the theory~\cite{Aoyama:2020ynm} and experiments, a new challenge has been put forth for the lattice community, which is to determine the hadronic contributions to the $(g-2)_\mu$ as precisely as possible. The $I=1$ $\pi\pi$ scattering study at the physical point plays a pivotal role in improving the precision of the lattice determination of $a_{\mu}^{\rm{hvp}}$, which is the leading order hadronic vacuum polarization contribution to the anomalous magnetic moment of the muon. It helps in improving the estimation of $a_{\mu}^{\rm{hvp}}$ at long distances in two ways: it reduces the statistical error to better than a percent level on the vector correlator by taking into account its overlap with the $\pi\pi$ states~\cite{gerardin:2019rua} and provides a quantitative estimate of the finite size corrections by computing the time-like pion form factor~\cite{Meyer:2011um, Bernecker:2011gh, Asmussen:2018oip, Feng:2014gba, Andersen:2018mau, Erben:2019nmx}. In the past few years, a number of groups~\cite{chakraborty:2017tqp, borsanyi:2017zdw, blum:2018mom, giusti:2019xct, shintani:2019wai, Davies:2019efs, gerardin:2019rua, Aubin:2019usy, Lehner:2020crt, Borsanyi:2020mff} in the community have made significant efforts in this direction. 

With new strides in algorithmic developments over the past decade, computation of $\rho(770)$ resonance parameters has become a benchmark study in the lattice community~\cite{Aoki:2007rd, Feng:2010es, Lang:2011mn, Pelissier:2012pi, Feng:2014gba, Wilson:2015dqa, Bali:2015gji, Guo:2016zos, Fu:2016itp, Alexandrou:2017mpi, Andersen:2018mau, Werner:2019hxc, Erben:2019nmx, Fischer:2020yvw}. At the physical point, $\rho(770)$ resonance lies above the inelastic threshold which makes it interesting in terms of the domain of applicability for the L\"uscher formalism~\cite{Luscher:1990ux, Peardon:2009gh}. Lattice computations for the scattering phase shift using $N_f=2$ simulations have reported a tension with experiments~\cite{Fischer:2020yvw}. In this talk, we present preliminary results of the spectrum and the corresponding phase shift obtained for the $I=1$ $\pi\pi$ amplitude using the E250 ensemble from the Coordinated Lattice Simulations (CLS) consortium with $m_\pi= 129.60(97)$ MeV and $N_f$ = $2 + 1$ dynamical fermions.
\section{Methodology}\label{method}
\textbf{Lattice Setup}:
We performed our measurements on the ensemble E250 which has been generated with a non-perturbatively $\mathcal{O}(a)$ improved Wilson fermion action and a tree-level $\mathcal{O}(a^2)$ improved L\"uscher-Weisz gauge action~\cite{Mohler:2020txx}. The valence quarks are implemented using the non-perturbatively improved Wilson-clover fermions. The lattice volume for E250 is $96^3\times192$ with a fine 
\begin{wrapfigure}{r}{0.5\textwidth}
    \includegraphics[width=0.5\textwidth]{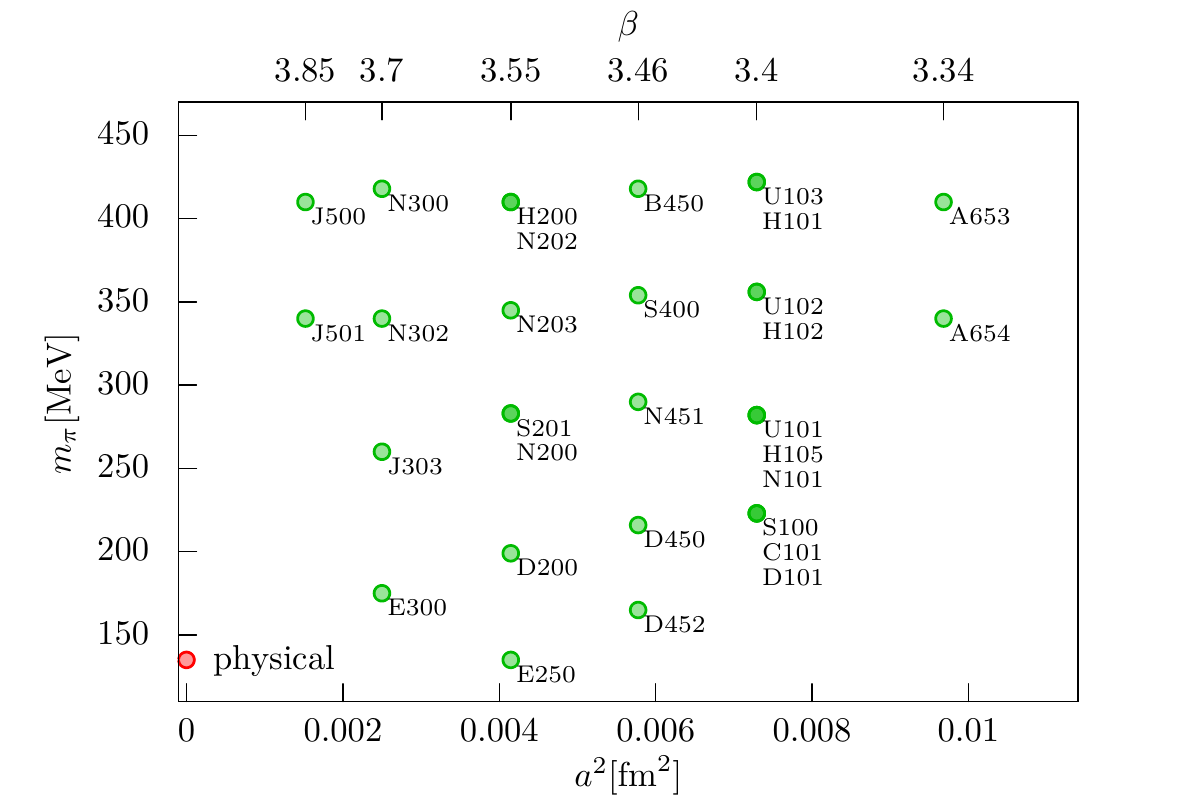}
    \caption{Different ensembles available within the CLS efforts. }
    \label{CLS_effort}
\end{wrapfigure}
lattice spacing of $a = 0.06426$ $\rm{fm}$ and periodic boundary conditions. This amounts to a physical box size of $L \approx 6.2$ $\rm{fm}$ which enables a momentum range of $0\,\leq\,|\vec{P}|^2\leq 4$ to map out the energy region upto $1$ GeV with sufficient granularity. A value of $m_\pi L = 4.1$ makes our measurements safe from any polarization effects. The source time-slices are distributed evenly for the periodic lattice. Analysis for another ensemble J303, close to the physical point at a finer lattice spacing~\ref{CLS_effort} is in the works and will be discussed in future publications. Calculations at two different lattice spacings will also help in constraining the finite lattice spacing effects.

\textbf{Distillation Setup}:
We employ the stochastic LapH method to treat all-to-all quark propagators~\cite{HadronSpectrum:2009krc, Morningstar:2011ka}. In the stochastic LapH framework, one computes the low modes of the three-dimensional gauge-covariant Laplacian and the quark sources are random vectors in the subspace spanned by these $N_{ev}$ eigenvectors. A stochastic estimate of the perambulator is obtained from $N_\eta$ noise sources, and we employ noise partitioning techniques \cite{Wilcox:1999ab,Foley:2005ac} to reduce the variance of the stochastic estimator. Two different types of quark lines have been evaluated to calculate the correlation functions. Quark lines emanating from a time slice and ending at all time slices of the lattice ({\lq}fixed lines{\rq}) are estimated employing full time dilution (denoted TF in \cite{Morningstar:2011ka}), while quark lines starting and ending on the same time slice ({\lq}relative lines{\rq}) are estimated using sources interlaced in time (TI). In addition we use interlacing in the Laplace eigenvectors index (LI) and full spin dilution (SF). This dilution scheme is utilized on $4$ evenly separated source time slices. The setup is detailed in Table \ref{tab:dilscheme}.  A total of $3072$ solutions of the Dirac equation per gauge configuration have been performed using the the \texttt{DFL\_SAP\_GCR} solver implemented in the \texttt{openQCD} package\footnote{\url{https://luscher.web.cern.ch/luscher/openQCD}}.   
\begin{table}
	\centering
	\begin{tabular}{c c c c c}
		\hline\hline
		quark-line type & dilution scheme & $t_0 /a$ & $N_\eta$ & $N_{ev}$\\
		\hline
		fixed & (TF,SF,LI16) & 4 random & 6 & 1536\\
		relative & (TI12,SF,LI16) & interlaced & 2 & 1536 \\
		\hline\hline
	\end{tabular}
	\caption{Dilution scheme, source times $t_0$, number of noise sources
		$N_\eta$ and number of Laplacian eigenvectors $N_{ev}$ used to estimate quark propagation in the computation of the spectrum on Ensemble E250.}
	\label{tab:dilscheme}
\end{table}
 In order to avoid potentially small eigenvalues during inversion, the gauge links are stout smeared before computing the eigenvectors with parameters ($\rho_{\rm{stout}}$, $N_{\rm{stout}}$) = ($0.1$, $36$). For this preliminary study, we have used $210$ gauge configurations which are well separated in the Monte Carlo chain to alleviate autocorrelations.  

\textbf{Interpolating operators}:
In our study, we have incorporated for the operator basis the following:

\begin{align}
\rho^+(\vec{P},t) &= \frac{1}{2 L^{3/2}} \sum_{\vec{x}} e^{-i \vec{P}\cdot \vec{x}} \bar{d} \Gamma u (t) \, ,
\end{align}
where $\Gamma  = \gamma_i(1\pm\gamma_4)$ and linear combinations of $\pi\pi$ operators of the form,  
\begin{align}
(\pi \pi)(\vec{p}_1,\vec{p}_2,t) =
\pi^+(\vec{p}_1,t)
\pi^0(\vec{p}_2,t)
-\pi^0(\vec{p}_1,t)
\pi^+(\vec{p}_2,t)
\, .
\end{align}
for various $\{\vec{p_1}, \vec{p_2}\}$ pairs to make them transform irreducibly. The momenta $\vec{p}_1$ and $\vec{p}_2$ of the single pions add
up to the frame momentum $\vec{P}$, i.e. $\vec{p}_1+\vec{p}_2=\vec{P}\equiv(2\pi/L)\vec{d}$, where $\vec{d}$ is a vector of integers. We will omit the prefactor $(2\pi/L)$ in $\vec{P}$ for convenience. The single-pion interpolators are defined by
\begin{align}
\pi^+ (\vec{q},t) =&
\frac{1}{2 L^{3/2}} \sum_{\vec{x}} e^{-i \vec{q} \cdot \vec{x}}
\big( \bar{u} \gamma_5 d \big) (\vec{x},t)
\, , \\
\pi^0 (\vec{q},t) =&
\frac{1}{2L^{3/2}} \sum_{\vec{x}} e^{-i \vec{q} \cdot \vec{x}}
\big( \bar{u}  \gamma_5 u - \bar{d}  \gamma_5 d \big) (\vec{x},t) \phantom{a} \, .
\end{align}
The correlation functions are computed using these interpolators in the isospin limit, after which they are analysed in 5 different frames with $0\,\leq\,|\vec{d}|^2\leq 4$. We employ the variational approach to disentangle the contributions to the states in the spectrum into the ground states and the relevant excited states. Thus, one constructs a correlation matrix from these correlation functions for each center-of-mass momentum and its irreducible representations. 

\textbf{Spectrum extraction}:
The variational method entails solving the Generalized EigenValue Problem (GEVP) for the correlation matrices\cite{Michael:1985ne}.
\begin{equation}
C_{ij}(t)v^{(n)}_j(t,t_0) = \lambda^{(n)}(t,t_0)C_{ij}(t_0)v^{(n)}_j(t,t_0).
\end{equation}
As $t\to\infty$ the eigenvalues $\lambda^{(n)}(t,t_0)$ have an asymptotic form $\lambda^{(n)}(t,t_0)\propto e^{-E_nt}$ where $E_n$ is the energy of the $n$-th eigen state that overlaps with our choice of interpolating operators. The GEVP is solved at $t = t_{eig}$ with a choice of $t_0$ such that the eigenvectors are insensitive to the variations in $t_0$ and $t_{eig}$. These normalized eigenvectors are then used to rotate the correlation matrices at other values of $t$ to obtain,
\begin{equation}
C_{\rm{rot}}^{(n)}(t) = \hat{v}^{(n)}_i(t,t_0)C_{ij}(t)\hat{v}^{(n)}_j(t,t_0).
\end{equation}
This method relies on the assumption that the eigenvectors are time independent for sufficiently large $t$. The domain of applicability of this method can be evaluated by measuring the magnitude of the off-diagonal elements in the rotated correlation matrices. We expect the $C_{\rm{rot}}^{(n)}(t)$ to be diagonal or approximately diagonal at most values of $t$.  

\begin{wrapfigure}{r}{0.42\textwidth}
 \centering
 \includegraphics[width=6.0cm]{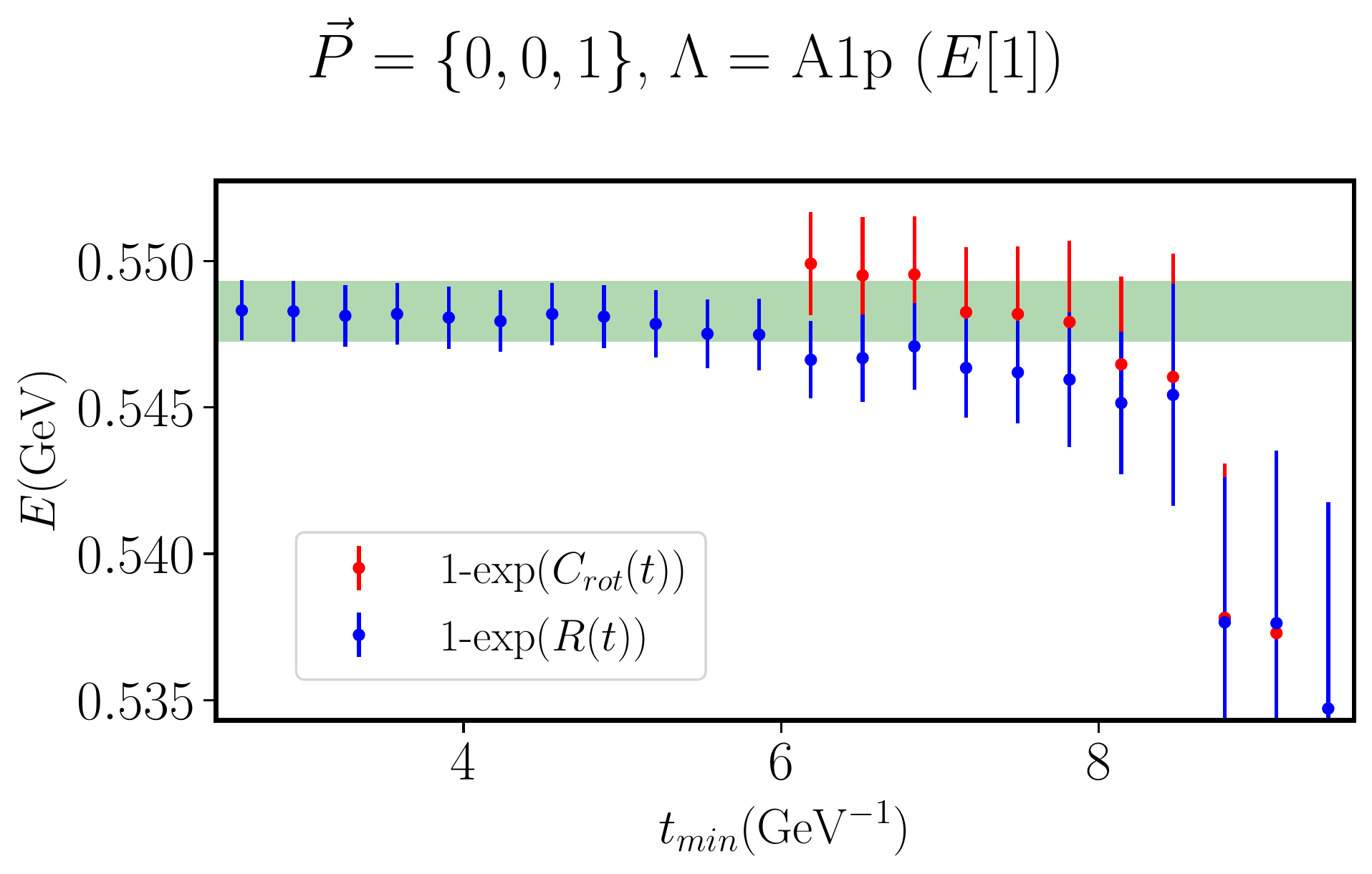}
 \caption{$t_{\rm{min}}$ plots with $t_{\rm{max}} = 10 $ $\rm{GeV}^{-1}$ to compare the reconstructed energy from ratio fits shown by blue points and the fits to the diagonal element of the $C_{\rm{rot}}^{(n)}(t)$ shown by red points, for the first excited state of the spectrum in the moving frame $\vec{P} = \{0,0,1\}$ and the irrep $\Lambda = A_1^+$. }
 \label{fitquality}
\end{wrapfigure}

The diagonal values of $C_{\rm{rot}}^{(n)}(t)$ are used to construct the ratio, 
\begin{equation}
R^{(n)}(t) = C_{\rm{rot}}^{(n)}(t)/(C_{\pi}^{\vec{p}_1}(t)C_{\pi}^{\vec{p}_2}(t)
\end{equation}
with a nearby non-interacting level [${\pi}^{\vec{p}_1}{\pi}^{\vec{p}_2}$]. The $R^{n}(t)$ quantifies the energy difference from that non-interacting level which is then fitted with a single exponential form to obtain the shift  $\Delta\mbox{E}=\ln(\frac{R^{(n)}(t)}{R^{(n)}(t+1)})$\cite{Bulava:2016mks} from the non-interacting level. The total energy can be reconstructed by adding the nearest $2$-pion non-interacting energy level to the $\Delta\mbox{E}$. In Figure  \ref{fitquality}, we present
   the reconstructed ${E}_{ \text{recon}} = (E_{\pi\pi}^{\vec{p}_1+\vec{p}_2} + \Delta\mbox{E})$ along with the energy obtained from the single exponential fits to the $C_{\rm{rot}}^{(n)}(t)$, for the first excited state in the irrep $A_1^+$~($+$ denotes $G$-parity) of the $|\vec{d}|^2=1$ moving frame on the E250 ensemble. The early occurrence of the plateau in the ratio fit is representative of the fact that it is effective in removing the contamination from higher excited states at earlier time slices whereas the eigenvalue correlator reaches that plateau at a much later time. For our preliminary study, our choices of $(t_{min},\, t_{max})$ are made using the ratio fits.

\textbf{Finite Volume Phase Shift Analysis:} In principle, L\"uscher quantization is not applicable beyond the $4\pi$ inelastic threshold because finite volume corrections to the Bethe-Salpeter kernel are no longer exponentially suppressed\cite{Luscher:1985dn, Luscher:1986pf, Luscher:1990ux}. However, the coupling to the inelastic thresholds below the $K\bar{K}$ threshold is extremely weak: the inelasticity parameter $\eta\geq 0.98$ below the $K\bar{K}$. Therefore, in our preliminary calculation, we neglect the $4\pi$ threshold~\cite{Bernard:2010fp, Giudice:2012tg} and employ the L\"uscher quantization condition to compute the scattering phase shifts in order to compare it with experiments. The L\"uscher quantization condition for elastic $\pi\pi$ scattering is
\begin{align}
\label{eq:QC}
{\rm det} \bigg( \mathbbm{1} + \I t_{\ell}(s)(\mathbbm{1} + \I {\cal M}^{\vec{P}}) \bigg) = 0\quad\text{where}\quad t_{\ell}(s) = \frac{1}{\cot{\delta_{\ell}(s)}-\I}.
\end{align}
The matrix ${\cal M}^{\vec{P}}$ has the indices ${\cal M}_{lm,l'm'}^{\vec{P}}$, where $l,l'$ label the irreducible representations of $SO(3)$ and $m,m'$ are the corresponding row indices. ${\cal M}^{\vec{P}}$  is a known finite-volume function. In the case of $P$-wave $\pi\pi$ scattering, different sectors of partial waves are approximately decoupled in the L\"uscher analysis and the contributions from $F$-wave and higher are highly suppressed~\cite{Dudek:2012xn}. Therefore, we discard the energy levels dominated by higher partial waves. 

\section{Results and Discussion}\label{results}
\begin{wrapfigure}{r}{0.45\textwidth}
    \vskip -0.8in
    \centering
    \subcaptionbox{}{\includegraphics[width=0.35\textwidth]{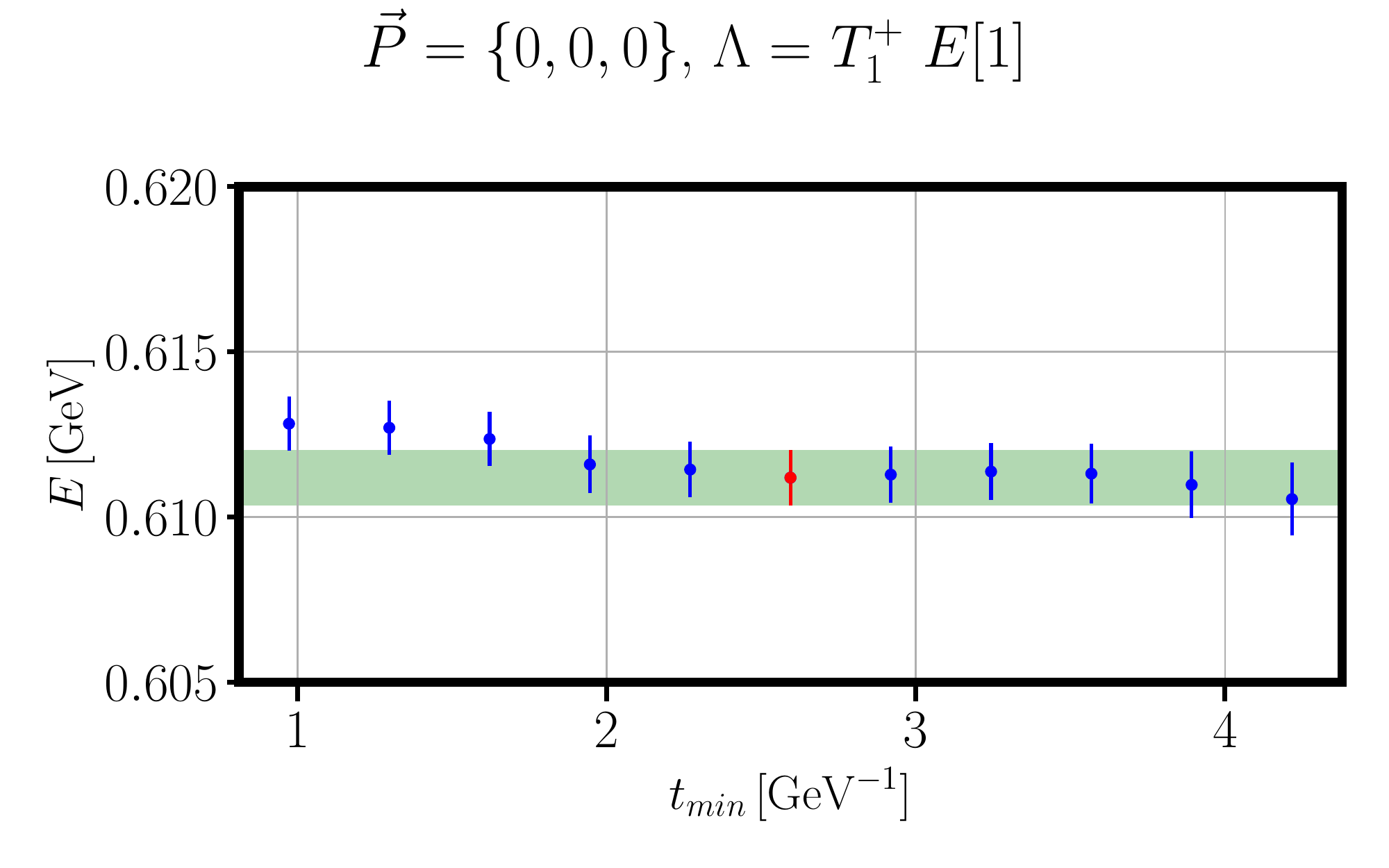}}\par
    \subcaptionbox{}{\includegraphics[width=0.35\textwidth]{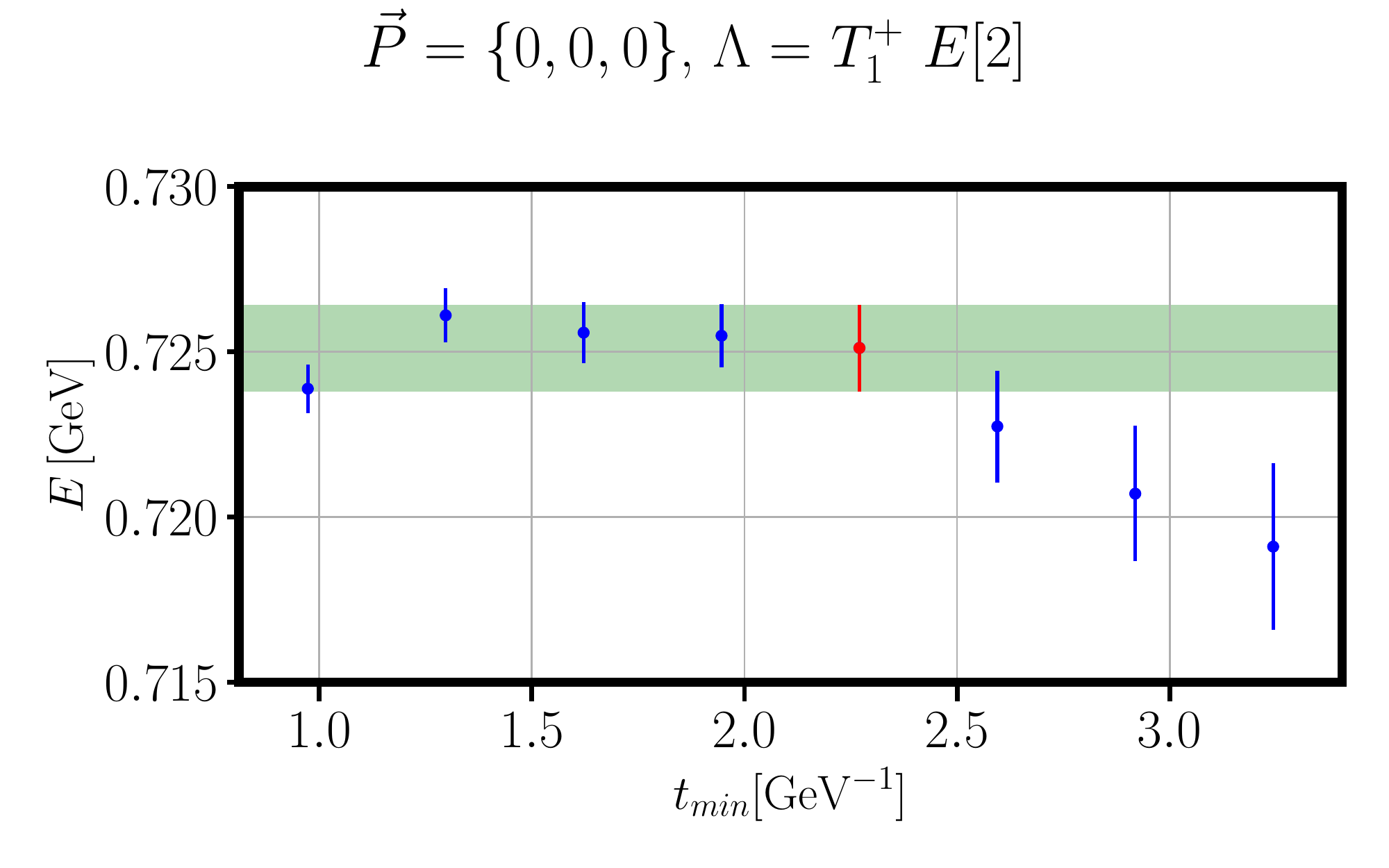}}\par
    \subcaptionbox{}{\includegraphics[width=0.35\textwidth]{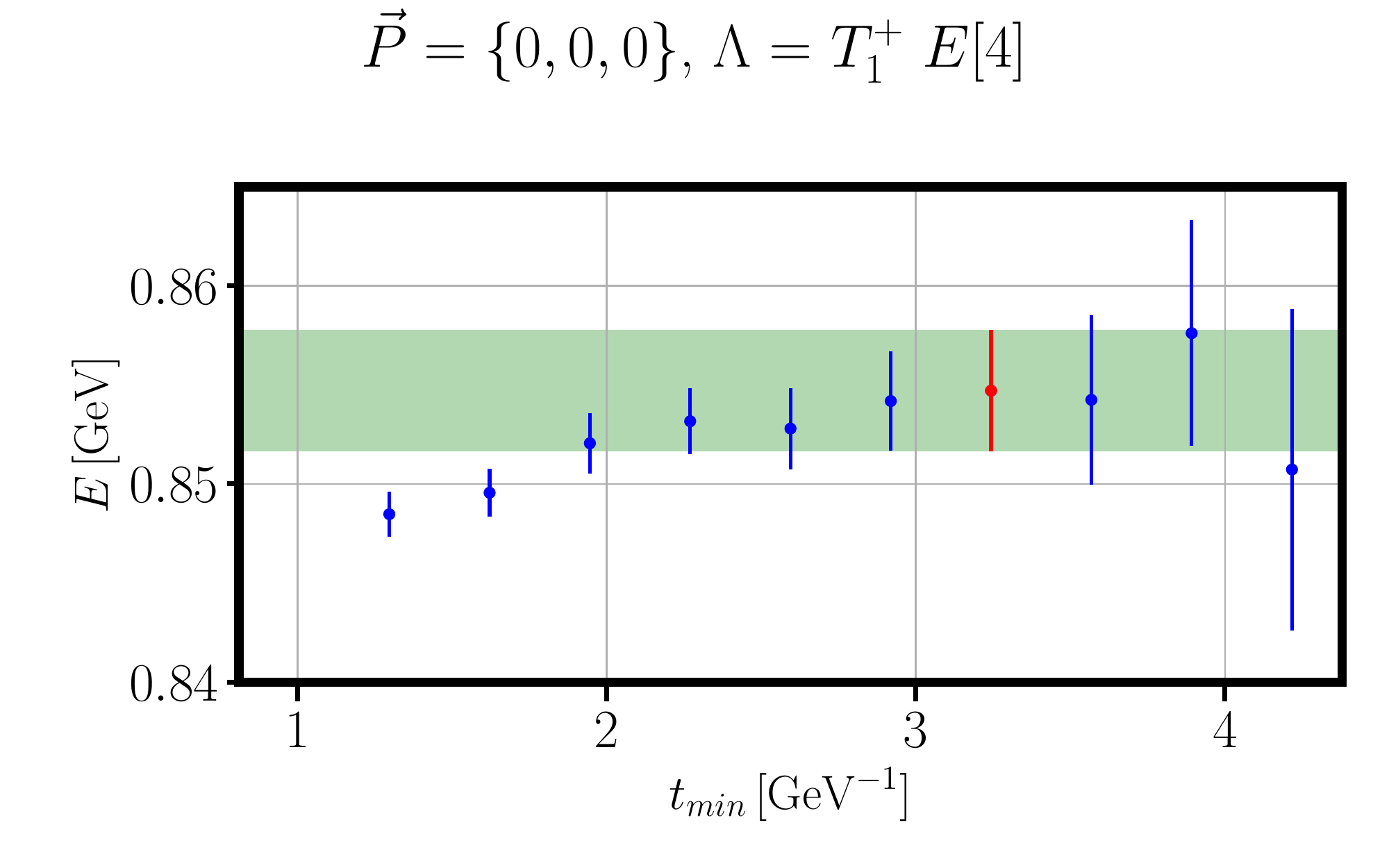}}
    \caption{Stability fit for the $1^{st}$, $2^{nd}$ and $4^{th}$ excited states for the spectrum of the $T_{1u}^+$ irrep in the CM frame where $t_{max} = 6.2$ $\rm{GeV}^{-1}$. The red point is the selected energy level.}
    
    \label{spectrum_150MeV}
\end{wrapfigure}
In Figures \ref{spectrum_150MeV}(a)-(c),  we show the preliminary fits to the finite-volume energy spectrum for the $T_{1u}^+$ irrep in the center-of-mass frame. At a first glance of Figures \ref{spectrum_150MeV} (a) to (c), we can infer that as we go from the first excited state to the higher excited states, the signal gets more noisy.
We note that in Figure \ref{spectrum_150MeV} (a) we are already above the inelastic $4\pi$ threshold which is $\approx 0.5$ GeV and we have good signal until $t_{min} = 4 \,\,\rm{GeV}^{-1}$. In the next Figure \ref{spectrum_150MeV}(b), as we approach the {\lq}physical{\rq} $\rho(770)$, the signal has a very early plateau, only until $2.5 \rm{GeV}^{-1}$. Subsequently, in Figure \ref{spectrum_150MeV}(c), as the spectrum drifts away from the {\lq}physical{\rq} $\rho(770)$, the signal gets noisier after around $4 \,\,\rm{GeV}^{-1}$. 

Qualitatively, we infer that the $t_{min}$ plateaus for the ratio fits shrink for a state whose overlap with the $\pi\pi$ interpolating operator is minimal, which justifies the inclusion of single hadron operators having the quantum numbers of the continuum $\rho$ resonance.

We need to ensure a stable $t_{min}$ variation of the spectrum because the L\"uscher quantization condition is highly non-linear. We have extracted the finite volume spectrum upto the $K\bar{K}$ threshold which is $\approx 1\,\,\rm{GeV}$. 

\begin{figure}
 \centering
 \includegraphics[width=12.0cm]{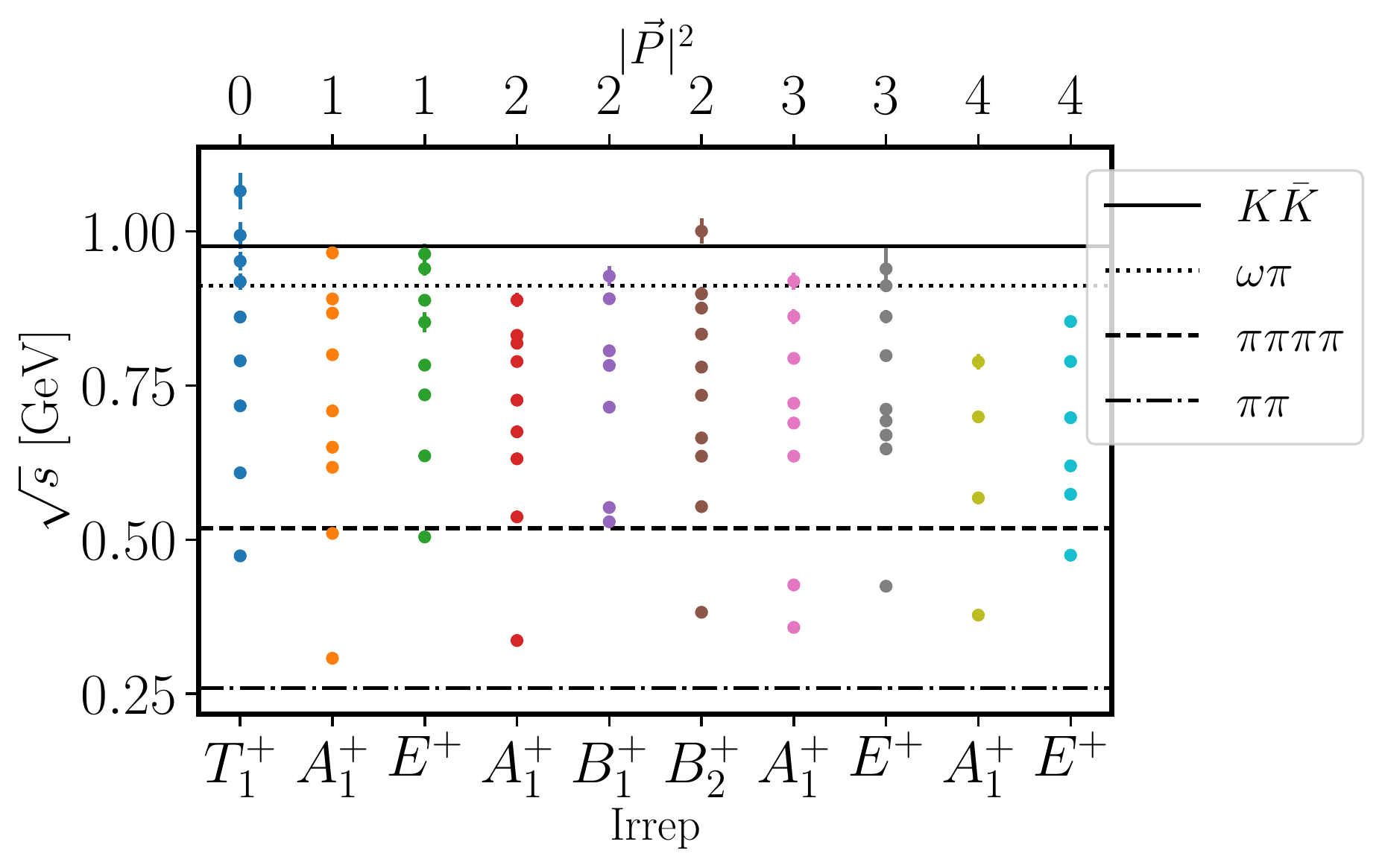}
 \caption{E250 spectrum for different irreps in different moving frames. }
 \label{full_spectrum_150MeV}
\end{figure}

The selected finite volume spectrum from the ratio fits for each irrep belonging to different moving frames have been plotted in Figure \ref{full_spectrum_150MeV}. All thresholds below the $K\bar{K}$ threshold have been depicted. We have $11$ energy levels below the $4\pi$ inelastic threshold and the rest of the levels, except $3$, are between the $\pi\pi$ and $K\bar{K}$ thresholds.  

\begin{figure}
 \centering
 \includegraphics[width=10.0cm]{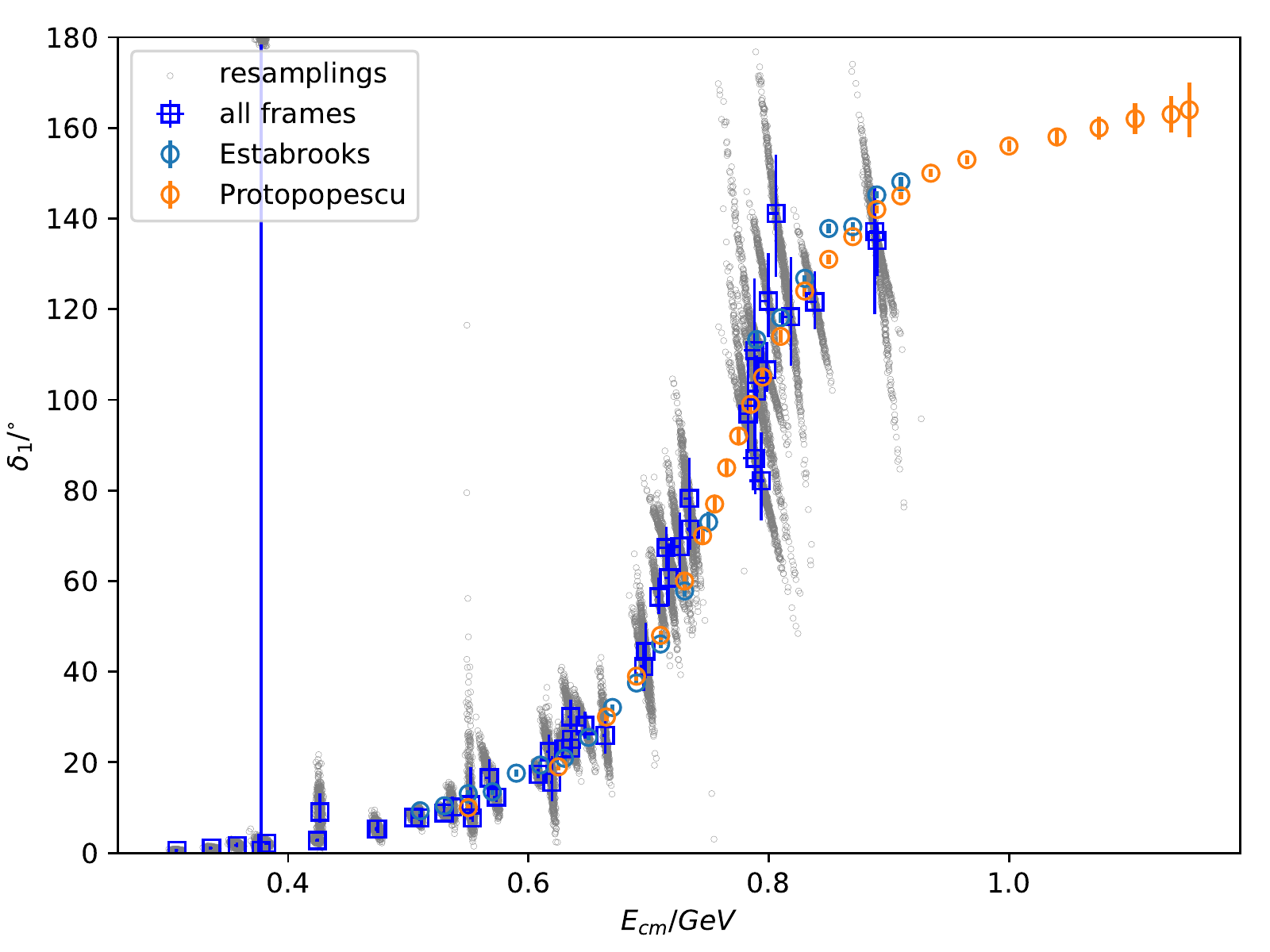}
 \caption{Phase shifts obtained from the lattice spectrum using L\"uscher quantization condition, along with the phase shifts obtained from experiment\cite{Protopopescu:1973sh, Estabrooks:1975cy}. The grey dots represent the phase shifts obtained on each bootstrap resampling.}
 \label{phase_shift}
\end{figure}
Utilizing the energy spectrum obtained from the lattice, we compute the scattering phase shifts for different center of mass energies as shown in Figure~\ref{phase_shift}. This preliminary result is showing no noticeable tension~\cite{Fischer:2020yvw} at this level of precision.
\section{Summary}
We report the status of $\pi\pi$ spectroscopy at the physical point pursued by the Mainz group, in order to aid the calculation of the $a_\mu^{\rm{hvp}}$ contribution to $(g-2)_\mu$. The analysis is performed on an $N_f=2+1$ CLS ensemble using the stochastic distillation framework. We discuss our preliminary estimate of the finite-volume spectra and the corresponding scattering phase shifts. The calculation of the time-like pion form factor is currently underway and efforts are being made to increase the statistics.
\acknowledgments
Calculations for this project used HPC resources on the supercomputers HAWK at High-Performance Computing Center(HLRS), Stuttgart. The authors gratefully acknowledge the support of the John von Neumann Institute for Computing and Gauss Centre for Supercomputing e.V.(\url{http://www.gauss-centre.eu}) for project GCS-HQCD. ADH is supported by: (i) The U.S. Department of Energy, Office of Science, Office of Nuclear Physics through the Contract No. DE-SC0012704 (S.M.); (ii) The U.S. Department of Energy, Office of Science, Office of Nuclear Physics and Office of Advanced Scientific Computing Research, within the framework of Scientific Discovery through Advance Computing (SciDAC) award Computing the Properties of Matter with Leadership Computing Resources. E250 gauge configurations were generated by DM within the CLS efforts on the supercomputers JUQUEEEN at Juelich Supercomputing Center, HAZEL HEN and HAWK at HLRS Stuttgart and MOGON II at Johannes Gutenberg University Mainz. We acknowledge computing resources provided through award 1913158
on Frontera at the Texas Advanced Computing Center (TACC). CJM acknowledges support from the U.S.~NSF under award PHY-1913158. D.M. acknowledges funding by the Heisenberg Programme of the Deutsche Forschungsgemeinschaft (DFG, German Research Foundation) – project number 454605793.

\bibliographystyle{JHEP}
\input{lattice2021.bbl}
\end{document}

%% file: macro.tex
\newcommand\bef{\begin{figure}}
\newcommand\eef[1]{\label{fg:#1}\end{figure}}
\newcommand\bec{\begin{center}}
\newcommand\eec{\end{center}}
\newcommand\besf{\begin{subfigure}}
\newcommand\eesf[1]{\label{sfg:#1}\end{subfigure}}
\newcommand\beq{\begin{equation}}
\newcommand\eeq[1]{\label{#1}\end{equation}}
\newcommand\beqa{\begin{eqnarray}}
\newcommand\eeqa[1]{\label{#1}\end{eqnarray}}
\newcommand\bet{\begin{table}}
\newcommand\eet[1]{\label{tb:#1}\end{table}}
\newcommand\best{\begin{subtable}}
\newcommand\eest[1]{\label{stb:#1}\end{subtable}}
\newcommand\betb{\begin{center}\begin{tabular}}
\newcommand\eetb{\end{tabular}\end{center}}
\newcommand\beit{\begin{itemize}}
\newcommand\eeit{\end{itemize}}

\newcommand{\cref}[1]{\bec  \textcolor{black}{\small #1}\eec}

\definecolor{DarkGreen}{rgb}{0.00,0.29,0.00}
\definecolor{DarkRedfooter}{rgb}{0.60,0.00,0.00}
\definecolor{DarkRed}{rgb}{0.60,0.00,0.00}
\definecolor{DarkRedtitle}{rgb}{0.85,0.00,0.00}

\def\prsp#1#2%
  {\mathop{}%
   \mathopen{\vphantom{#2}}^{#1}%
   \kern-\scriptspace%
   #2}

%% file: lattice2021.bbl
\providecommand{\href}[2]{#2}\begingroup\raggedright\endgroup